\documentclass[aps,prl,amsmath,amssymb,10pt,superscriptaddress,twocolumn,color,epsfig,graphicx,bm]{revtex4-1}

\usepackage{color}
\usepackage{graphicx}
\usepackage{graphics}
\usepackage{subfigure}
\usepackage{epsfig}
\usepackage{amsmath}
\usepackage{array}
\usepackage{amssymb}
\usepackage{graphicx}      % Include figure files
\usepackage{dcolumn}       % Align table columns on decimal point
\usepackage{bm}            % bold math
\usepackage{units}
\usepackage{array}
   
\newcommand{\ub}{Department of Physics, Universit\"{a}t Basel,
Klingelbergstr. 82, 4056 Basel, Switzerland}
\newcommand{\lpmcn}{Universit\'e de Lyon, F-69000 Lyon, France and
LPMCN, CNRS, UMR 5586, Universit\'e Lyon 1, F-69622 Villeurbanne, France}
\begin{document}

\title{Conducting boron sheets formed by the reconstruction of the $\alpha$-boron (111) surface}

\author{Maximilian Amsler}
\affiliation{\ub}
\author{Silvana Botti}
\affiliation{\lpmcn}
\author{Miguel A.L. Marques}
\affiliation{\lpmcn}
\author{Stefan Goedecker}
\email{Stefan.Goedecker@unibas.ch}
\affiliation{\ub}

\date{\today}

\begin{abstract}
% The reconstruction of the $\alpha$-boron (111) surface was extensively
% studied based on \textit{ab initio} calculations in combination with a
% systematic structural search employing the minima hopping
% method. 
Systematic \textit{ab initio} structure prediction was applied 
for the first time to predict low energy surface
reconstructions by employing the minima hopping
method on the $\alpha$-boron (111) surface.
Novel reconstruction geometries were identified and carefully
characterized in terms of structural and electronic properties. Our
calculations predict the formation of a planar, mono-layer sheet at
the surface, which is responsible for conductive surface
states. Furthermore, the isolated boron sheet is shown to 
be the ground state 2D-structure in vacuum at a hole density of
$\eta=1/5$ and is therefore a potential
candidate as a precursor for boron nano-structures.
%Furthermore, the isolated boron sheet is shown to be
%thermodynamically stable in vacuum and is therefore a potential
%candidate as a precursor for boron nano-structures.

\end{abstract}

\pacs{somepacs}

\maketitle

Boron exhibits an impressive variety in forming chemical bonds due to
its trivalent electronic configuration, ranging from covalent
2-electron 2-center (2e2c) bonds to polycentered,
metallic-like~\cite{albert_boron:_2009} as well as ionic
~\cite{oganov_ionic_2009,mondal_electron-deficient_2011} bonding,
resulting in a high structural diversity.  In the solid state and at
ambient conditions, the existence of at least two polymorphs is widely
accepted: the low-temperature $\alpha$-rombohedral boron phase
($\alpha$-B)~\cite{mccarty_new_1958} and the high temperature
modification of $\beta$-rombohedral boron
($\beta$-B)~\cite{sands_rhombohedral_1957,hughes_structure_1963}. Several
other polymorphs have been reported, such as the high-pressure
modification $\gamma$-B$_{28}$~\cite{oganov_ionic_2009}, or the
much-disputed tetragonal boron I (t-I) and tetragonal boron II (t-II),
also called
$\alpha$-tetragonal~\cite{laubengayer_boron._1943,hoard_structure_1951,hoard_structure_1958}
and $\beta$-tetragonal
boron~\cite{talley_new_1960,vlasse_crystal_1979}. The main structural
motif of all above phases is the formation of interlinked B$_{12}$
icosahedra, a common characteristic of many other boron rich
compounds which can further contain various different
triangular-defined polyhedral building
blocks~\cite{albert_boron:_2009}.

Recently, immense efforts have been made in studying boron structures
of lower dimensionality. Theoretical predictions of boron
nanotubes~\cite{boustani_nanotubules_1997,gindulyte_proposed_1998,boustani_new_1999}
with metallic conductivity independent of helicity have drawn
significant attention in search for one dimensional
nanostructures. From the experimental point of view there have been
reports on mono- and multi-walled boron
nanotubes~\cite{ciuparu_synthesis_2004,liu_metal-like_2010}, boron
nanowires~\cite{cao_well-aligned_2001,otten_crystalline_2002,tian_boron_2008,liu_fabrication_2008,tian_patterned_2009},
boron nanorods~\cite{zhu_synthesis_2009}, nano
belts~\cite{wang_catalyst-free_2003} and
nanoribbons~\cite{xu_crystalline_2004} which may be used for future
applications in nano-electronics. In analogy to carbon nanotubes
formed from graphene, boron nanotubes have been theoretically studied
by rolling two dimensional boron sheets~\cite{quandt_boron_2005}. In
contrast to graphene, boron sheets have can be either
buckled~\cite{kunstmann_approach_2007}, consisting of a triangular
lattice, or, according to more recent predictions, as planar
structures with a partially filled honeycomb structure at a specific
hole density $\eta$, the latter being energetically
favorable~\cite{tang_novel_2007,tang_first-principles_2010}. Due to the structural
flexibility and the competing energetic ordering the exact structural
and compositional ground state is still under discussion,
leading to numerous studies to identify the most stable 2D lattice
configuration in vacuum~\cite{yu_prediction_2012,wu_two-dimensional_2012,lu_binary_2013}
and on substrates~\cite{liu_probing_2013} at different hole densities. 
Similarly, again inspired by their carbon
counterparts, various hollow molecular structures have been
theoretically
proposed~\cite{marques2005,gonzalez_szwacki_b_80_2007,botti2009,ozdogan_unusually_2010,quarles_filled_2011,pochet_low-energy_2011,li_b80_2012}
in analogy to the C$_{60}$ and other carbon
fullerenes~\cite{kroto_c60:_1985}.

However, there is surprisingly little work on boron at the
intermediate dimension between bulk and 2D sheets, namely on surfaces
of boron and boron nano crystals. Experimental measurements
approximate the surface energy of $\alpha$-B to be in the range of
$173-401$~meV/\AA$^2$ from cracks in boron
fibers~\cite{vega-boggio_surface_1977}, and early theoretical studies
have been carried out based on empirical potential models resulting in
282~meV/\AA$^2$ for the (111) surface of the same
polymorph~\cite{vega-boggio_calculation_1977}. \textit{Ab initio}
calculations were carried out by W. Hayami to systematically estimate
the energies of low index surfaces in the $\alpha$-B, $\beta$-B, t-I
and t-II phases~\cite{hayami_surface_2007,hayami_role_2007}. However,
none of the above studies have taken into account any reconstruction
mechanisms which are in fact frequently observed in most
semi-conducting materials. Clearly, reconstructions can lead to considerable decrease in
surface energy, thereby significantly altering surface reactivity and
other properties.

With increasing computational resources it has become popular to
predict structures of crystalline materials and clusters, and more
recently of 2D planar structures by means of sophisticated global
optimization algorithms~\cite{oganov_2010}. The present work however
goes one step further by applying a structure prediction 
scheme directly on surfaces to study reconstructions of large boron
surfaces with the minima hopping method
(MHM)~\cite{goedecker_2004,amsler_2010} based on \textit{ab initio}
calculations. 
%ADDED
To our knowledge no studies on \textit{ab initio} 
predictions of surface 
reconstructions have been reported in literature 
without some sort of input from experiment.
%ADDED
The MHM was designed to predict energetically favorable
structures by exploring the potential energy surface with a
combination of consecutive short molecular dynamic trial steps
followed by local geometry optimizations. Many earlier applications
have shown the predictive power of the
MHM~\cite{amsler_crystal_2012,flores-livas_high-pressure_2012,amsler_novel_2012},
including investigations on surface structures of atomic force
microscopy silicon model tips~\cite{ghasemi_2008,amsler_2009}.

Globally optimizing large surface slabs with more than hundred atoms
is computationally prohibitive when performed at the density
functional theory (DFT) level. Therefore, initial MHM simulations were
conducted within the density functional based tight binding method, an
approximate DFT scheme, as implemented in the DFTB+
package~\cite{Aradi_DFTB+_2007}, to roughly map out the energy surface
and produce a variety of low energy structures, which were
subsequently fed back into the MHM using more accurate DFT
calculations to refine the search. The projector augmented wave
formalism was employed as implemented in
VASP~\cite{kresse_efficient_1996} using the Perdew-Burke-Erzernhof
(PBE)~\cite{PBE96} exchange-correlation functional which has been
shown to give highly reliable energy differences between different
structural motifs in boron~\cite{hsing_quantum_2009}. Final results
were refined with a plane-wave cut-off energy of 500~eV and
sufficiently dense k-point meshes such that the total energy was
converged to better than 1~meV per atom. Four additional
exchange-correlation-functionals were employed to confirm the
energetic ordering of the lowest lying structures, namely
PBEsol~\cite{perdew_restoring_2008}, the local density approximation
(LDA), the HSE06 hybrid functional~\cite{heyd_hybrid_2003,paier_erratum:_2006,heyd_erratum:_2006} as well as
BLYP~\cite{becke_density-functional_1988,lee_development_1988} within
the {\sc abinit} plane wave DFT
code~\cite{gonze_brief_2005,bottin_large-scale_2008}. Geometries were
fully relaxed with a tight convergence criteria of less than
0.004~eV/\AA~for the maximal force components.

\begin{figure}[h]            % Figure 1 : Structures
 \includegraphics[width=0.8\columnwidth,angle=0]{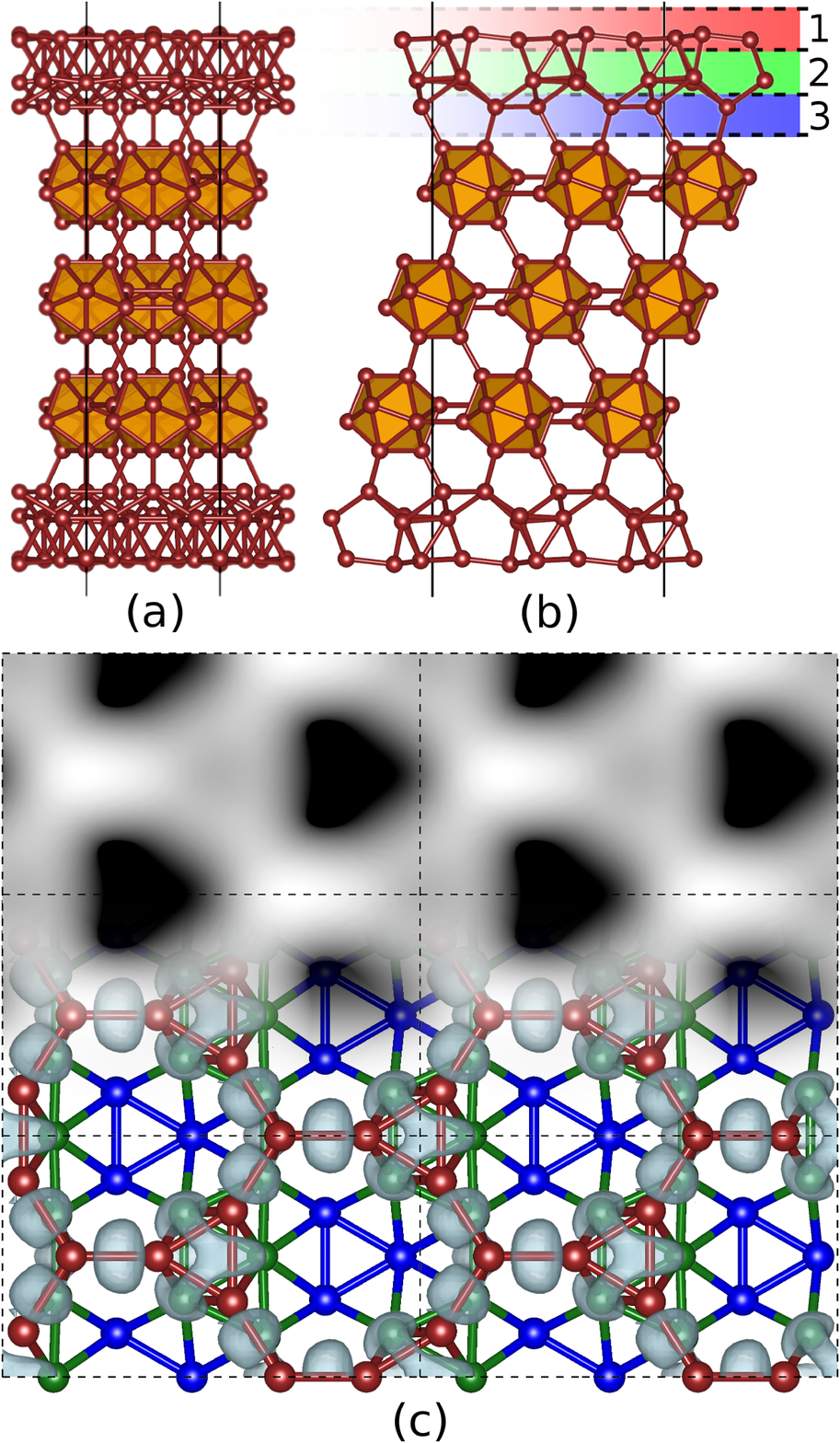}
  \caption{(Color online) View from the $\langle 1 1 \overline{2} \rangle $ (a) and the $\langle 1 \overline{1} 0 \rangle$ (b) direction of (111)-I$_{\mathrm{R,(h)}}$. (c) Top view towards the surface from the $\langle \overline{1} \overline{1} \overline{1}\rangle$ direction. The colors red, green and blue denote the first, second and third atomic layer, respectively. An isosurface of the ELF is shown for the top two layers at the bottom, and the STM image is overlayed at the top.}
  \label{fig:108}
\end{figure}

$\alpha$-B is described by cubic close packing of B$_{12}$ units,
while planes of icosahedra are stacked in $ABC$ order along the
$\langle111\rangle$ direction (see Fig.~\ref{fig:108} (a) and (b)
where the bulk layers are depicted by orange icosahedra). The MHM
simulations were conducted on slabs with up to 4-layers of B$_{12}$
units while at least all atoms in the topmost icosahedral layer were
allowed to move during the search. The majority of our simulations
were performed on a super-cell
$\{\mathbf{a}_\mathrm{S},\mathbf{b}_\mathrm{S},\mathbf{c}_\mathrm{S}\}$
generated from the rhombohedral cell vectors
$\{\mathbf{a}_\mathrm{R},\mathbf{b}_\mathrm{R},\mathbf{c}_\mathrm{R}\}$,
containing $\approx100$ atoms and such that a surface area of
62.6~\AA$^2$ was covered:
$\mathbf{a}_\mathrm{S}=2\mathbf{a}_\mathrm{R}-\mathbf{b}_\mathrm{R}-\mathbf{c}_\mathrm{R}$,
$\mathbf{b}_\mathrm{S}=\mathbf{a}_\mathrm{R}+\mathbf{b}_\mathrm{R}-2\mathbf{c}_\mathrm{R}$
and $\mathbf{c}_\mathrm{S}=\alpha
(\mathbf{a}_\mathrm{R}+\mathbf{b}_\mathrm{R}+\mathbf{c}_\mathrm{R})$,
for $\alpha>1$. To prevent interactions with periodic images along the
surface-normal direction, initial searches were conducted with a
vacuum layer of $\approx6$~\AA. The final results were obtained with a
vacuum space of more than 7~\AA, a value at which the surface energy
was converged better than 1~meV/\AA$^2$. Surface energies were computed according to
$\sigma=\frac{1}{2A}(N\epsilon_{\text{bulk}}-E_{\text{slab}})$, where 
$A$ denotes the surface area, $\epsilon_{\text{bulk}}$ is the 
energy per atom in the crystalline phase (here $\alpha$-B) and 
$E_{\text{slab}}$ is the energy of the slab containing $N$ atoms.
% was sufficiently converged.

Earlier investigations by Hayami~\textit{et
  al.}~\cite{hayami_role_2007} on unreconstructed (111) surfaces
indicated a difference in surface energy depending on whether the
surface was formed by inter- or intra-icosahedral cuts. According to
their results obtained with BLYP, a surface obtained by cutting
between icosahedral planes, denoted as (111)-I, would destroy the
presumably strong inter-icosahedral bonds and thus lead to a higher
surface energy compared to performing a cut through the weaker
intra-icosahedral bonds, denoted as (111)-II. However, when
reconstructions are permitted, this simple characterization based on
bond-breaking can easily fail. In fact, the number of mobile surface
atoms is much more important since it can lead to unexpected
structural rearrangements and the formation of new, energetically
favorable bonds. Therefore, MHM simulations were performed on both the
(111)-I and (111)-II surfaces and the results were summarized in
table~\ref{tab:energies}. Strikingly, the energetic difference of the
unreconstructed (111)-I and (111)-II surfaces depend strongly on the
employed exchange-correlation functional (first section in
table~\ref{tab:energies}). While PBE, PBEsol, LDA and HSE06 are in good
agreement and predict a small energy difference of $3-5$~meV/\AA$^2$
between (111)-I and (111)-II, BLYP gives a higher value of
$40-50$~meV/\AA$^2$, suggesting that the difference between inter- and
intra-icosahedral bond energies is less prominent than assumed by
Hayami~\textit{et al.}~\cite{hayami_role_2007}. The lowest energies of
the reconstructed surfaces (111)-I$_{\mathrm{R,(a)}}$ and
(111)-II$_{\mathrm{R,(a)}}$ from the MHM simulations are also
summarized in table~\ref{tab:energies} and clearly show that
(111)-I$_{\mathrm{R,(a)}}$ is energetically favored, in contrast to
the unreconstructed counterparts.

% Table generated by Excel2LaTeX from sheet 'Tabelle1'
\begin{table}[h]
%  \centering
  \caption{Lowest energy surfaces, sorted with respect to surface types. The surface energies are given for different functionals where available in units of meV/\AA$^2$. The lowest reconstructed structures (111)-I/II$_{\mathrm{R},(\alpha)}$ are given in the second section, enumerated  $\alpha=\mathrm{a, b,\dots}$ according to the PBE energetic ordering. Reconstructed surfaces with 3 additional and 3 missing atoms per super cell are denoted as (111)-I$^{+}_{\mathrm{R}}$ and (111)-I$^{-}_{\mathrm{R}}$, respectively.}
\begin{ruledtabular}
\label{tab:energies}
\begin{tabular} {l c c c c c }
Surface & PBE & PBEsol & LDA & HSE\footnotemark[1] & BLYP \\
    \hline
\multicolumn{5}{l}{Non-reconstructed} &\\
(111)-I   & 218.8& 233.3 &236.5& 247.5 & 205.4\\
  &&&&& 220.0\footnotemark[2] \\
(111)-II  & 215.6 &227.2 &231.7& 250.2 & 167.1\\
  &&&&& 170.0\footnotemark[2]\\
    \hline
\multicolumn{4}{l}{Reconstructed (111)-I} &\\
(111)-I$_{\mathrm{R,(a)}}$  & 170.6& 181.6& 190.4& 196.3   & $-$ \\ 
(111)-I$_{\mathrm{R,(b)}}$  & 172.2& 182.9& 192.3& 198.6   & $-$ \\
(111)-I$_{\mathrm{R,(c)}}$  & 172.5& 183.3& 192.4& 198.0   & $-$  \\
(111)-I$_{\mathrm{R,(d)}}$  & 173.1& 184.1& 192.7& 198.4   & $-$  \\ 
(111)-I$_{\mathrm{R,(e)}}$  & 173.6& 184.6& 192.8& 198.4   & $-$  \\ 
(111)-I$_{\mathrm{R,(f)}}$  & 173.8& 185.0& 193.2& 199.7   & $-$  \\ 
(111)-I$_{\mathrm{R,(g)}}$  & 173.8& 184.8& 193.3& 199.2   & $-$  \\ 
(111)-I$_{\mathrm{R,(h)}}$  & 174.3& 185.6& 192.5& 198.7   & $-$  \\ 
(111)-I$_{\mathrm{R,(i)}}$  & 175.4& 186.4& 194.8& 199.8   & $-$  \\ 
    \hline
\multicolumn{5}{l}{Reconstructed (111)-II} &\\
(111)-II$_{\mathrm{R,(a)}}$  & 183.0 & 200.8 & 206.6 & 210.0 & $-$\\
    \hline
\multicolumn{5}{l}{Reconstructed (111)-I, atoms added}  &\\
(111)-I$^{+}_{\mathrm{R,(a)}}$ & 199.4 & 212.4 & 217.9 & 226.1 & $-$ \\
    \hline
\multicolumn{5}{l}{Reconstructed (111)-I, atoms removed}& \\
(111)-I$^{-}_{\mathrm{R,(a)}}$   &180.4 & 194.6 & 201.3& 204.6& $-$  \\
\end{tabular}
\end{ruledtabular} 
\footnotetext[1]{Structure relaxed with PBE}
\footnotetext[2]{Taken from Ref.~\cite{hayami_role_2007}}
\end{table}

\begin{figure}[h]            % Figure 1 : Structures
\includegraphics[width=0.9\columnwidth,angle=0]{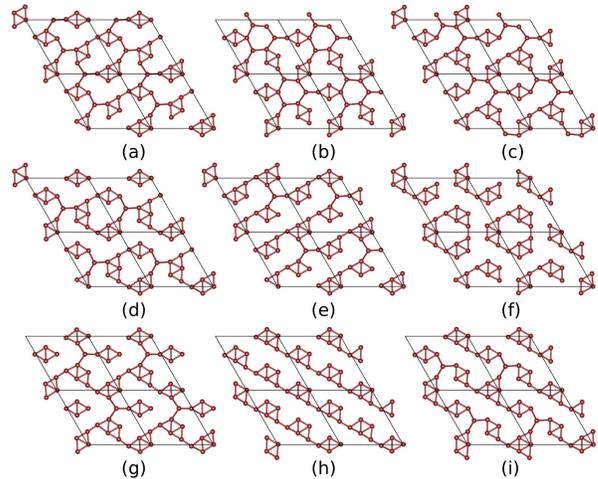}
   \caption{(Color online) The second atomic layer of the 9 lowest energy surface structures (111)-I$_{\mathrm{R},(\alpha)}$, $\alpha=\mathrm{a, b,\dots}$, are shown.}
   \label{fig:2ndlayer}
 \end{figure}

The arrangement of the atoms in the structure
(111)-I$_{\mathrm{R,(h)}}$ are shown in Fig.~\ref{fig:108}, where
panel (a) and (b) represent the side views along the
$\langle11\bar{2}\rangle$ and the $\langle1\bar{1}0\rangle$ direction,
respectively. The three outer atomic layers participate in the
reconstruction and are shown separately in panel (c). The first atomic
layer, colored in red, forms an almost planar filament of interlinked
triangular pattern with $Cmmm$ symmetry. It is identical for all low
energy structures that were found during the MHM simulations, thus
providing the fundamental structure to terminate the surface. This
first layer is supported by a second atomic layer of higher complexity
comprised by a denser network of triangular units (green), thereby
forming octahedra together with the triangles from the first
layer. The structural variety in this layer however is much larger as
illustrated in Fig.~\ref{fig:2ndlayer}. The corresponding energies are
given in the second section of table~\ref{tab:energies} which show
that the energetic differences resulting from the subtle
rearrangements of the triangles are very small, preventing an
unambiguous identification of the ground state structure. Finally, the
third layer is again identical for all low-energy structures (shown in
blue) and are simply the remaining base planes of the underlying
icosahedral bulk structure.

In contrast to (111)-I$_{\mathrm{R}}$, the reconstructed
(111)-II$_{\mathrm{R}}$ surfaces do not form any planar
structures. Since only half of the atoms are available for
reconstruction due to the intra-icosahedral truncation, the number of
mobile surface atoms is insufficient to satisfy an energetically
optimal bonding configuration. Although an additional layer of
icosahedra was explicitly included to rearrange during the MHM
simulation, the atoms in this layer did not participate in the
reconstruction and retained the bulk arrangement. The basic structural
motif for all low-energy (111)-II$_{\mathrm{R}}$ surfaces essentially
consists of loosely connected nano-ribbons with triangular pattern on
top of a perfect icosahedra layer of the bulk slab.

The compositional stability of the surface reconstructions was further
investigated by adding and subtracting atoms to the surface layers and
performing further MHM simulations (the last two sections of
table~\ref{tab:energies}). When depositing atoms (up to 3 atoms per
super cell) on the reconstructed surface no structural changes were
observed both in the second and third atomic surface layers. The
additional atoms clustered on the top layer by forming trigonal
structures oriented away from the surface. The surface energy of the
lowest structure of this type, (111)-I$^{+}_{\mathrm{R,(a)}}$, was
found to be 28.74~meV/\AA$^2$ higher in energy than the lowest
reconstructed structure (111)-I$_{\mathrm{R,(a)}}$. Similarly,
removing atoms from the surface layer (up to 3 atoms per super cell)
resulted in the reordering of the topmost layer without significantly
disturbing the second and third atomic layers. In fact, in the case of
removing 3 atoms per super cell, the top layer reconstructed such that
a triangular unit was missing, thus destroying an octahedron initially
formed together with the second layer. Although still endothermic, the
energy difference of (111)-I$^{-}_{\mathrm{R,(a)}}$ with respect to
the clean slab (111)-I$_{\mathrm{R,(a)}}$ is smaller with a value of
9.75~meV/\AA$^2$. It can therefore be concluded that the stoichiometry
obtained by a planar cut through the bulk is indeed the most stable.

Finally, the electronic structure was studied by investigating the
electron localization function (ELF)~\cite{silvi_classification_1994}
as implemented in VASP. Fig.~\ref{fig:108}~(c) shows a corresponding
ELF isosurface of the two top atomic surface layers. A clear
distinction can be made between 2e2c bonds in the topmost layer
connecting the triangular units in the plane, and the polycentric
bonds which dominate within all triangular arrangements, especially in
the octahedra formed between the first and the second layer. This
mixture of covalent 2e2c and polycentric bonding system is
characteristic also in the bulk region, where the intra-icosahedral
bonds are purely polycentric and inter-icosahedral bonds are either
2-centered (between planes) or 3-centered (within the
plane). Additionally, a constant current scanning tunnel microscopy 
(STM) simulation was performed at a bias of $V_b=-1.5$~eV based on 
the Tersoff-Hamann approximation~\cite{tersoff_theory_1985}. The constant 
current image at a maximal height of $\approx 1.8$~\AA~above the surface was 
rendered on top of Fig.~\ref{fig:108}~(c) which can be qualitatively 
compared to STM experiments .

\begin{figure}[t]
\setlength{\unitlength}{1cm}
\includegraphics[width=0.9\columnwidth,angle=0]{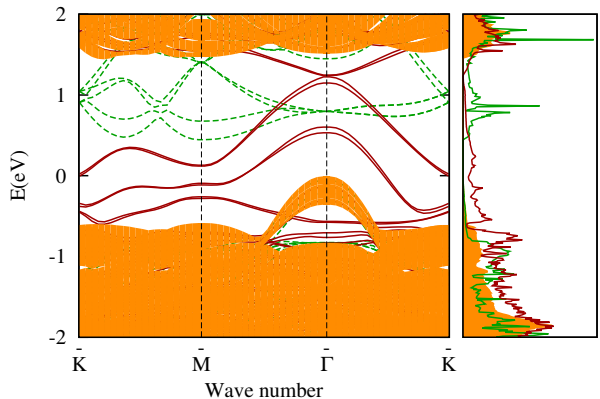}
\caption{The electronic surface band structures and the density of
  states of the (111)-I surface are shown. Orange lines (shaded)
  indicate bulk bands, whereas the red lines represent the surface
  states of the reconstructed surface. The green lines represent the
  surface bands of the unreconstructed surface of $\alpha$-B. The
  energy zero is set at the Fermi energy of the bulk.}
\label{fig:bands}
\end{figure}

\begin{figure}[t]
\setlength{\unitlength}{1cm}
\includegraphics[width=0.8\columnwidth,angle=0]{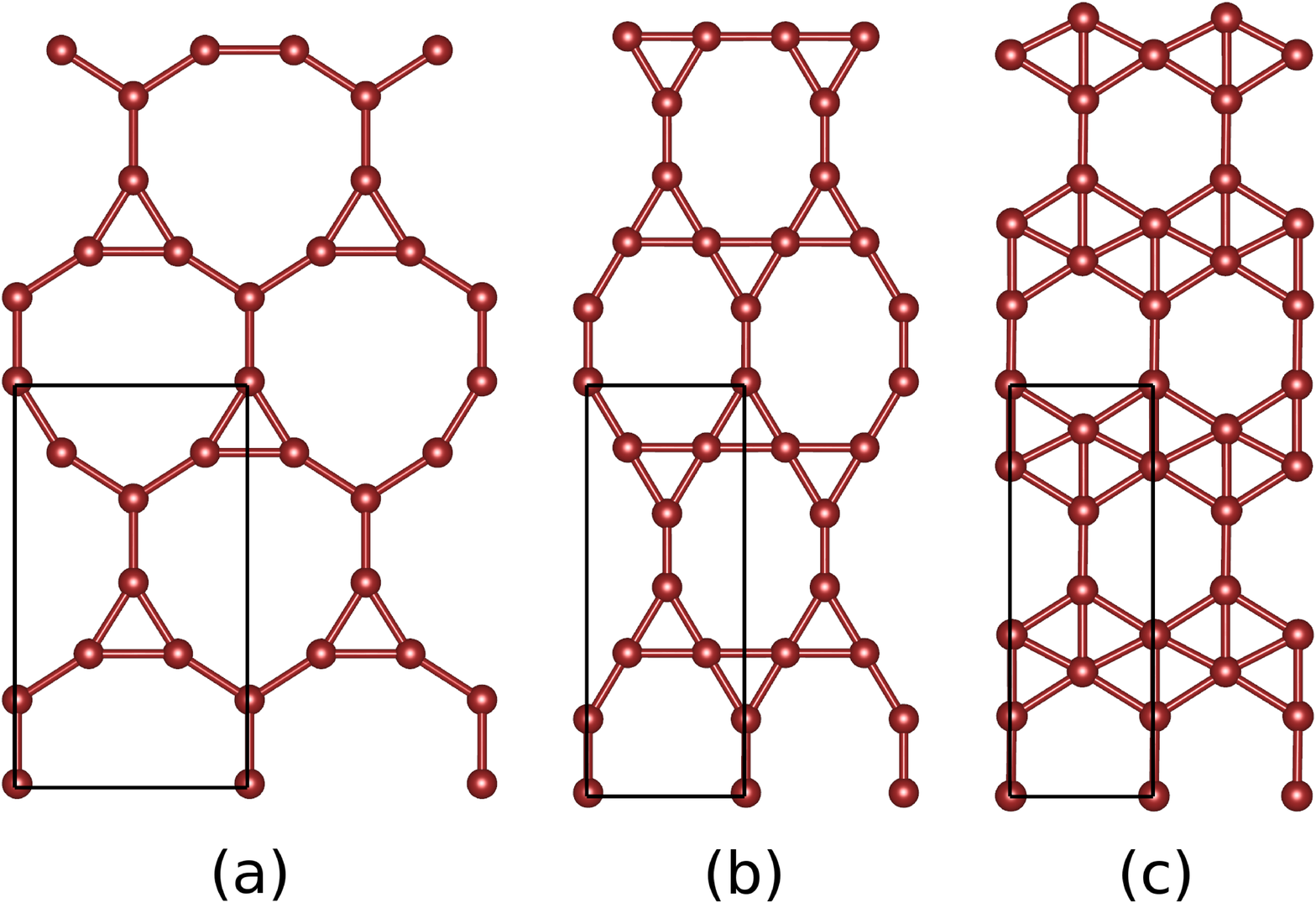}
\caption{Structural transformation of the topmost atomic layer of the
reconstructed surface during a local geometry optimization. (a) Initial structure, (b) intermediate during relaxation, (c) final structure, relaxed into the global minimum of 2D sheets at $\eta=1/5$ hole density}
\label{fig:monolayer}
\end{figure}

Furthermore, the electronic surface band structure was calculated
within the 2D-Brillouin zone for both the reconstructed and the
non-reconstructed surface as shown in Fig.~\ref{fig:bands}. The
projected band structure of bulk $\alpha$-B, shown as the shaded
region in orange, has a PBE band gap with a value of 1.47~eV. When a
vacuum is introduced into the material without further reconstruction,
the conduction bands decrease and appear as surface states in the gap
region which reduce its magnitude to 0.50~eV (shown as dashed, green
lines). However, conducting surface states emerge upon allowing the
surface to reconstruct, crossing the Fermi level and thus forming a
metallic surface (shown as solid, red lines). Since conventional DFT
commonly underestimates the band gap additional calculations were
performed with the HSE06 hybrid
functional,
leading to a bulk gap of 2.02~eV and 0.86~eV for the unreconstructed
(111)-I.  Earlier theoretical works have predicted many 2D boron
sheets to be metallic as
well~\cite{yu_prediction_2012,wu_two-dimensional_2012,lu_binary_2013}
such that one way of interpreting the above results is as if a planar,
conducting sheet of boron is adsorbed on a (111) $\alpha$-B
surface. Based on this perception the stability of the topmost atomic
layer was investigated when isolated from the surface. A simple local
geometry relaxation was performed in vacuum while allowing the cell to
adjust according to the in-plane stresses, starting from the initial
structure shown in Fig.~\ref{fig:monolayer}~(a). As expected, this
structure is metallic and transformed to a more compact state upon
relaxation to compensate the missing interaction with the
substrate. The final structure is shown in
Fig.~\ref{fig:monolayer}~(c), which has previously been predicted to
be the ground state in boron sheets at a fixed hole density of
$\eta=1/5$, denoted as ``struc-1/5'' in
Ref.~\cite{yu_prediction_2012}. 
%ADDED
No imaginary phonon frequencies
were found in the whole 2D Brillouin zone, indicating that
the 2D structure is dynamically stable. 
The energy difference between (111)-I$_{\mathrm{R,(a)}}$ and
the isolated sheet plus the optimized, remaining slab is merely 54.4~meV/\AA$^2$.
%ADDED
This sheet configuration is again metallic
and could provide a promising precursor for other boron
nano-structures such as single-walled boron nano-tubes or
cage-molecules.

In conclusion, an extensive, systematic \textit{ab initio} 
structural search was for the first time performed with the MHM on
surfaces to identify reconstruction geometries. 
The main structural motif of the reconstructed (111)
surface of $\alpha$-B was isolated, studied and accurately characterized.
In contrast to the unreconstructed
surface (111)-I, which merely reduces the band gap of crystalline
$\alpha$-B, the reconstructed (111)-I$_{\mathrm{R}}$ structures
contains metallic surface states that facilitates electric
conduction. The topmost layers of all low energy reconstructions
(111)-I$_{\mathrm{R}}$ are planar and consist of a 2D network of
interlinked triangular patterns. These results can be interpreted as
if a conducting sheet of boron was adsorbed on a semi-conducting
substrate, leading to numerous possible applications in
nano-electronics. Assuming that it is further possible to isolate the
top mono-layer of boron through exfoliation it would provide a 2D
boron sheet with $\eta=1/5$ hole density as a promising precursor for
a large variety of boron nano-structures.

Financial support provided by the Swiss National Science Foundation and French ANR (ANR-08-CEXC8-008-01 and ANR-12-BS04-0001-02)
are gratefully acknowledged. Computational resources were provided by the Swiss National Supercomputing
Center (CSCS) in Manno and French GENCI (project x2013096017).

%% \bibliography{Bib_Boron}

\end{document}